\documentstyle[12pt,wrapft,epsbox,twoside,fleqn,espcrc1]{article}

\newcommand{\AmS}{{\protect\the\textfont2
  A\kern-.1667em\lower.5ex\hbox{M}\kern-.125emS}}

\title{Analyses of collective flow and space-time evolution 
based on relativistic hydrodynamical model }

\author{C. Nonaka\address{Department of Physics, Hiroshima University, 
        Higashi-hiroshima,\\ Hiroshima 739-8526, Japan} 
        , N. Sasaki$^{\rm a}$ , S. Muroya\address{Tokuyama Women's College,
        Tokuyama, Yamaguchi 745-8511, Japan } 
        and   
        O. Miyamura$^{\rm a}$}

\begin{document}
\maketitle

\begin{abstract}
We numerically solve fully (3+1)-dimensional relativistic 
hydrodynamical equation with the baryon number conservation law.
For realistic initial conditions we adopt the 
results from the event generator (URASiMA). 
Using this model we discuss collective flow.
\end{abstract}

\section{INTRODUCTION}
The various kinds of flow phenomena such as directed flow, 
elliptic flow and radial flow have been observed in recent 
experiment at AGS and SPS \cite{E877f}-\cite{WA98}.
Anisotropic flow phenomena are expected to show a strong dependence 
on the nuclear 
equation of state \cite{Danielewicz}. Here,  
for analyzing anisotropic flow phenomena with relativistic 
hydrodynamical model in detail, we need {\it full} calculation 
of (3+1)-dimensional relativistic hydrodynamical equation.
Our purpose is analysis of collective flow based on (3+1)-dimensional 
relativistic hydrodynamical model with realistic initial conditions.

\section{MODEL DESCRIPTION}
\begin{wrapfigure}{r}{55mm}
\epsfile{file=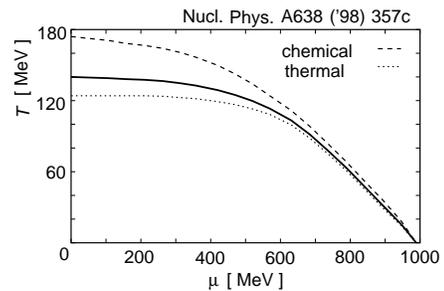,scale=0.25}
\caption{The solid line stands for freeze-out in the 
present calculation. 
The dashed line and the dotted line are referenced in \cite{Heinz}.}
\end{wrapfigure}
Our calculational procedure is explained briefly.
To begin with, we calculate initial conditions for the relativistic 
hydrodynamical model by using the event generator, URASiMA
(Ultra-Relativistic A-A collision
Simulator based on Multiple scattering Algorithm) which  
is characterized by Multi chain model \cite{Date}.  
We focus on AGS energy region and several incident energies such as 
5, 10.8, 20 AGeV for Au+Au collisions in which 
impact parameter is fixed to 6 fm \cite{E895}. 
We characterize time $t_{0}$ when the projectile nuclei 
finishes passing through the target nuclei and hydrodynamical 
evolution starts.
Energy density and baryon density from URASiMA is 
transformed into temperature and chemical potential with the equation of 
state of the ideal hadron gas.
Here, so as to solve the relativistic 
hydrodynamical equation easily we smooth the results keeping 
condition that 
the difference between original results and 
smoothed results is less than 1 \%.
We numerically solve fully (3+1)-dimensional relativistic
hydrodynamical equation coupled with the baryon number conservation 
law using Lagrangian hydrodynamics \cite{Ishii}.
The equation of state is needed to solve the hydrodynamical 
equation and anisotropic flow is evaluated in detail 
by inputting the different equation of state to the 
hydrodynamical equation.   
Here, for the first trial we use the equation of state of the ideal 
hadron gas which contains resonance up to 2 GeV \cite{Particle-data}.
Furthermore, 
we assume that the hadronization process occurs when the temperature and
chemical potential of the volume elements cross the boundary (solid
line in fig.1). The solid line is so designed that the freeze-out temperature
becomes 140 MeV at vanishing chemical potential, based on chemical
freeze-out model and thermal freeze-out model \cite{Heinz}.
Finally, we investigate azimuthal distribution of proton by using 
Cooper-Frye Formula \cite{Cooper-Frye} for particle emission 
from hadronic fluid.

\section{CALCULATED RESULTS}
\begin{figure*}[htb]
\begin{minipage}[t]{52mm}
\begin{center}
$t=5$ [fm/c]
\end{center}
\epsfile{file=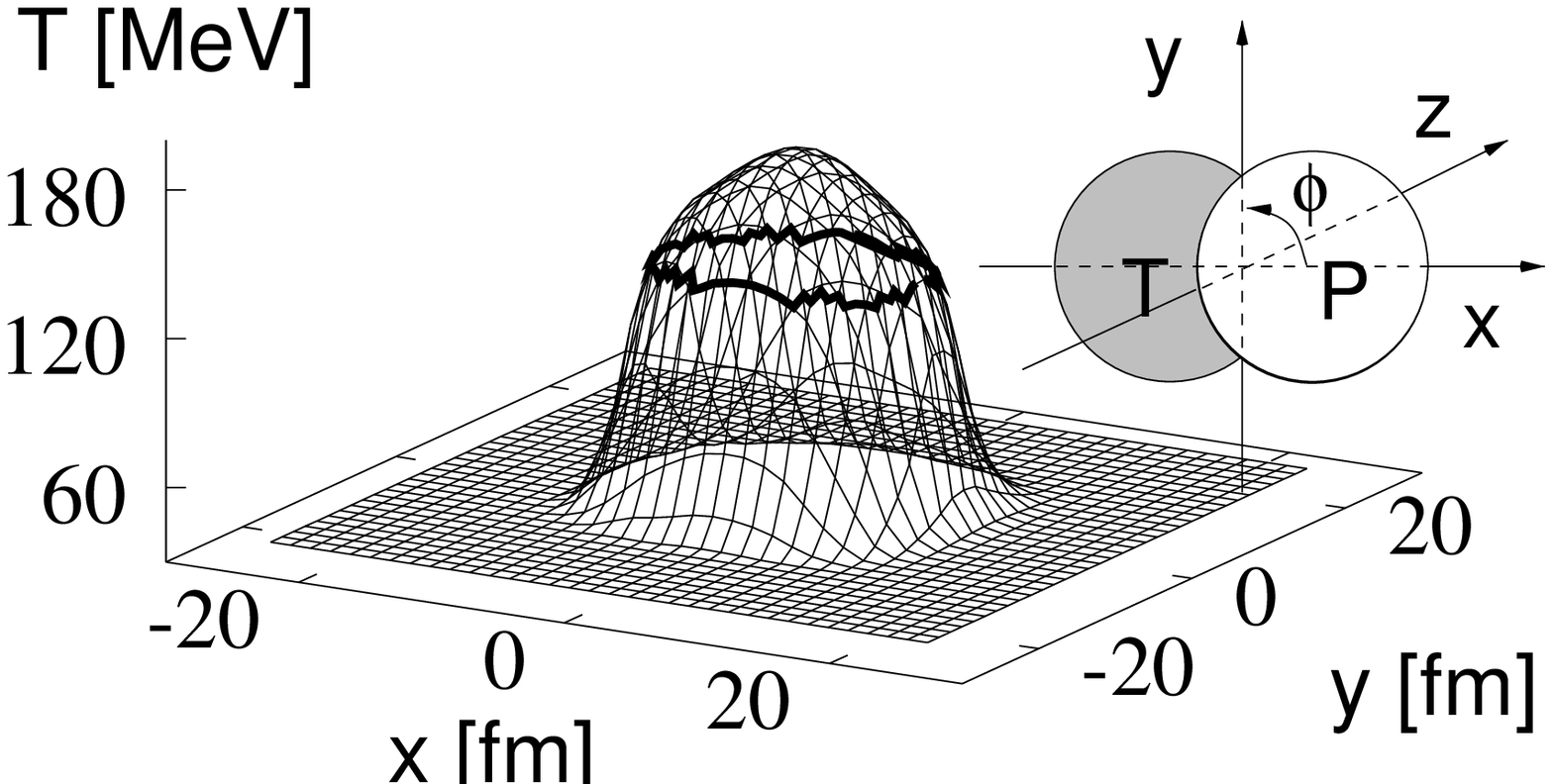,scale=0.29}
\end{minipage}
\begin{minipage}[t]{52mm}
\begin{center}
$t=15$ [fm/c]
\end{center}
\epsfile{file=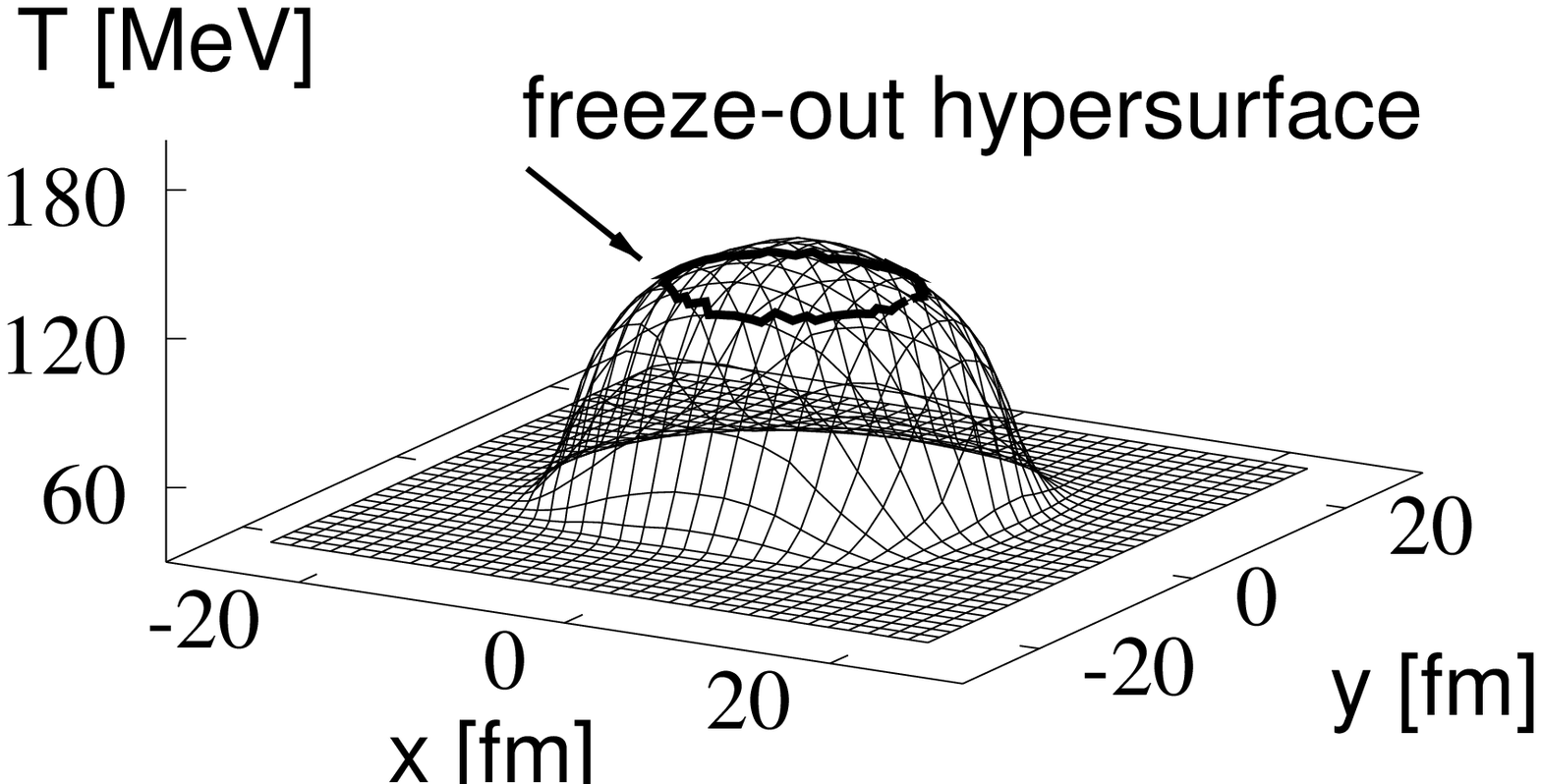,scale=0.29}
\end{minipage}
\begin{minipage}[t]{52mm}
\begin{center}
$t=25$ [fm/c]
\end{center}
\epsfile{file=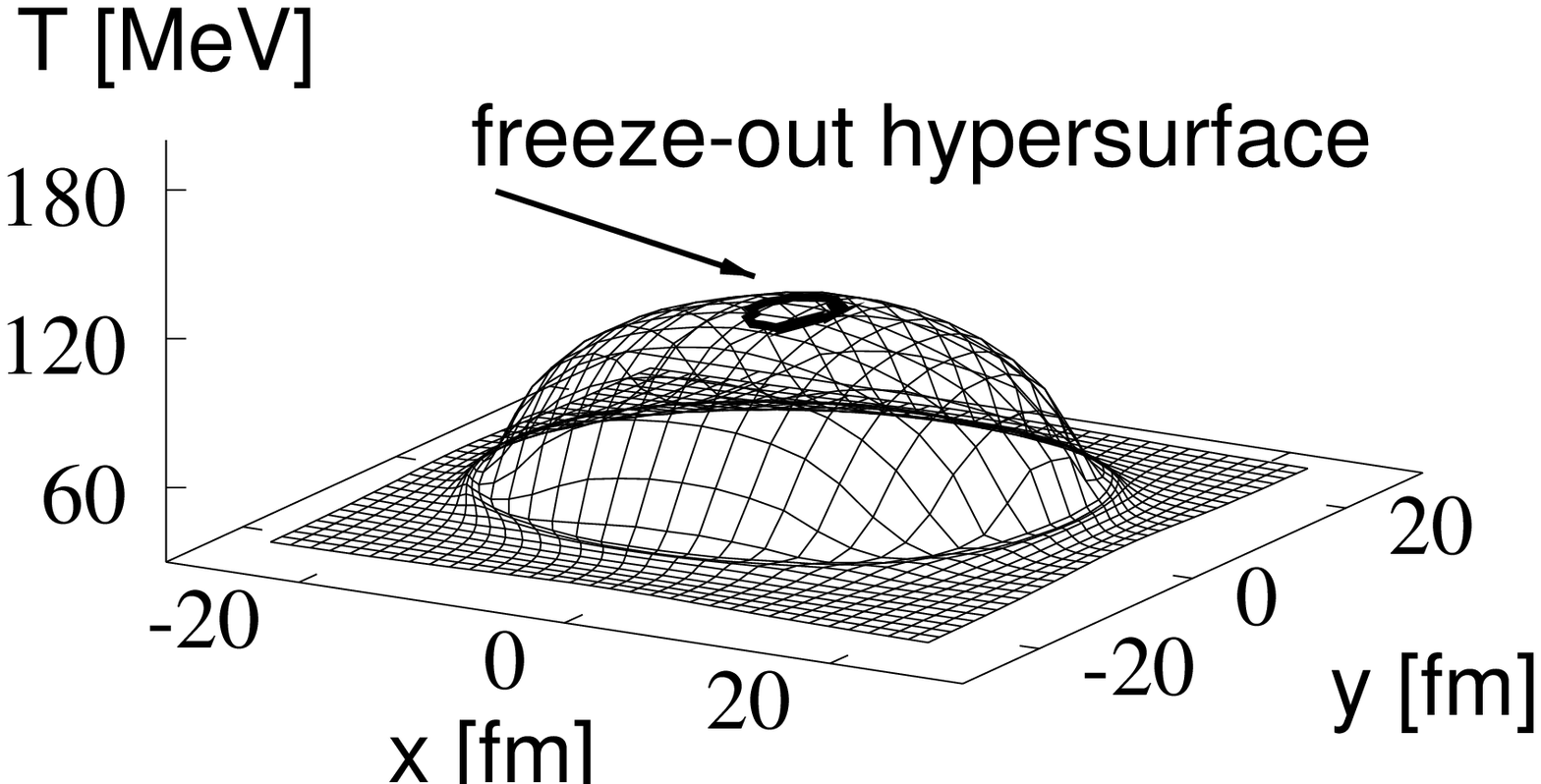,scale=0.29}
\end{minipage}
\begin{minipage}[t]{52mm}
\epsfile{file=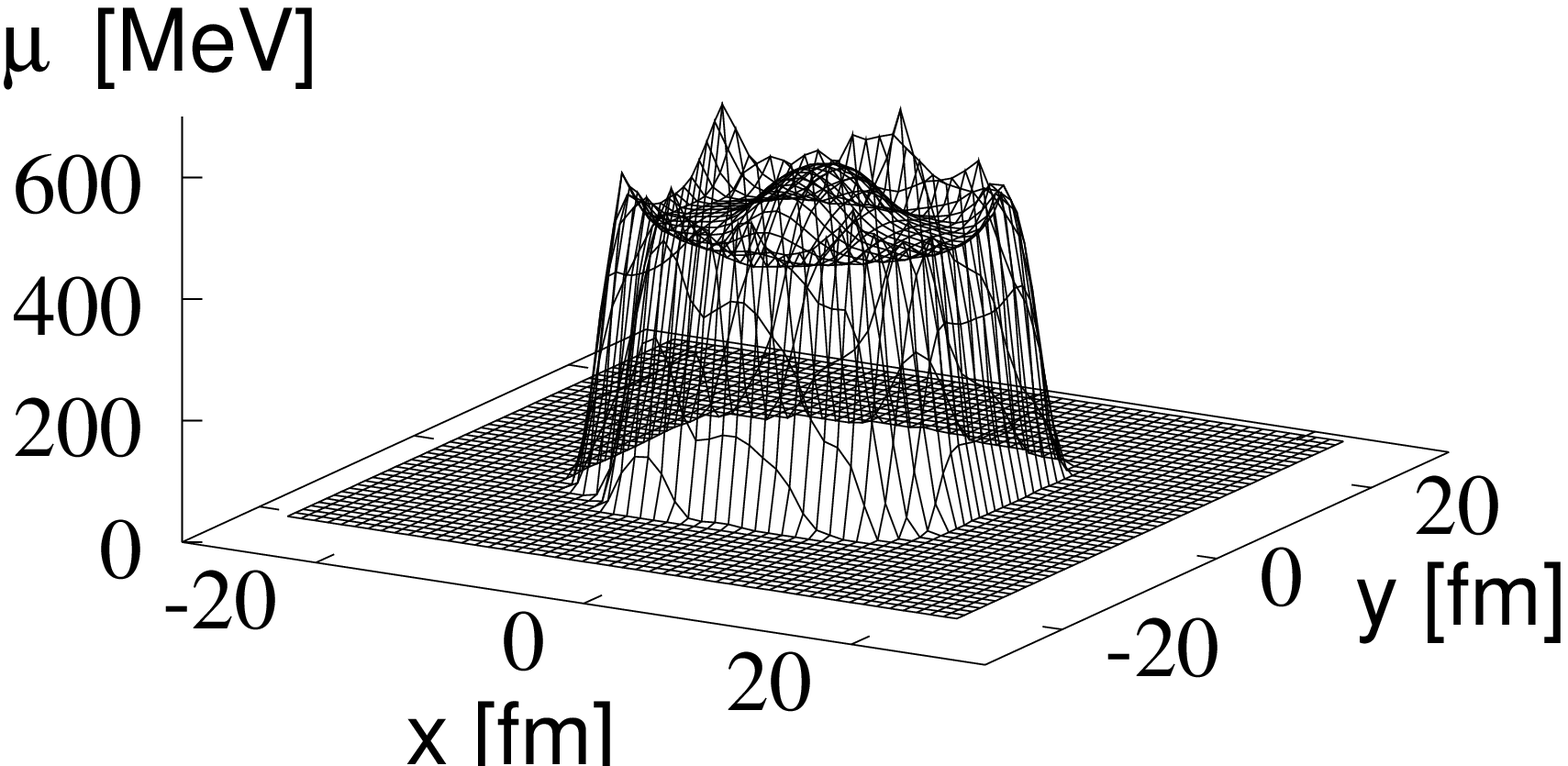,scale=0.29}
\end{minipage}
\begin{minipage}[t]{52mm}
\epsfile{file=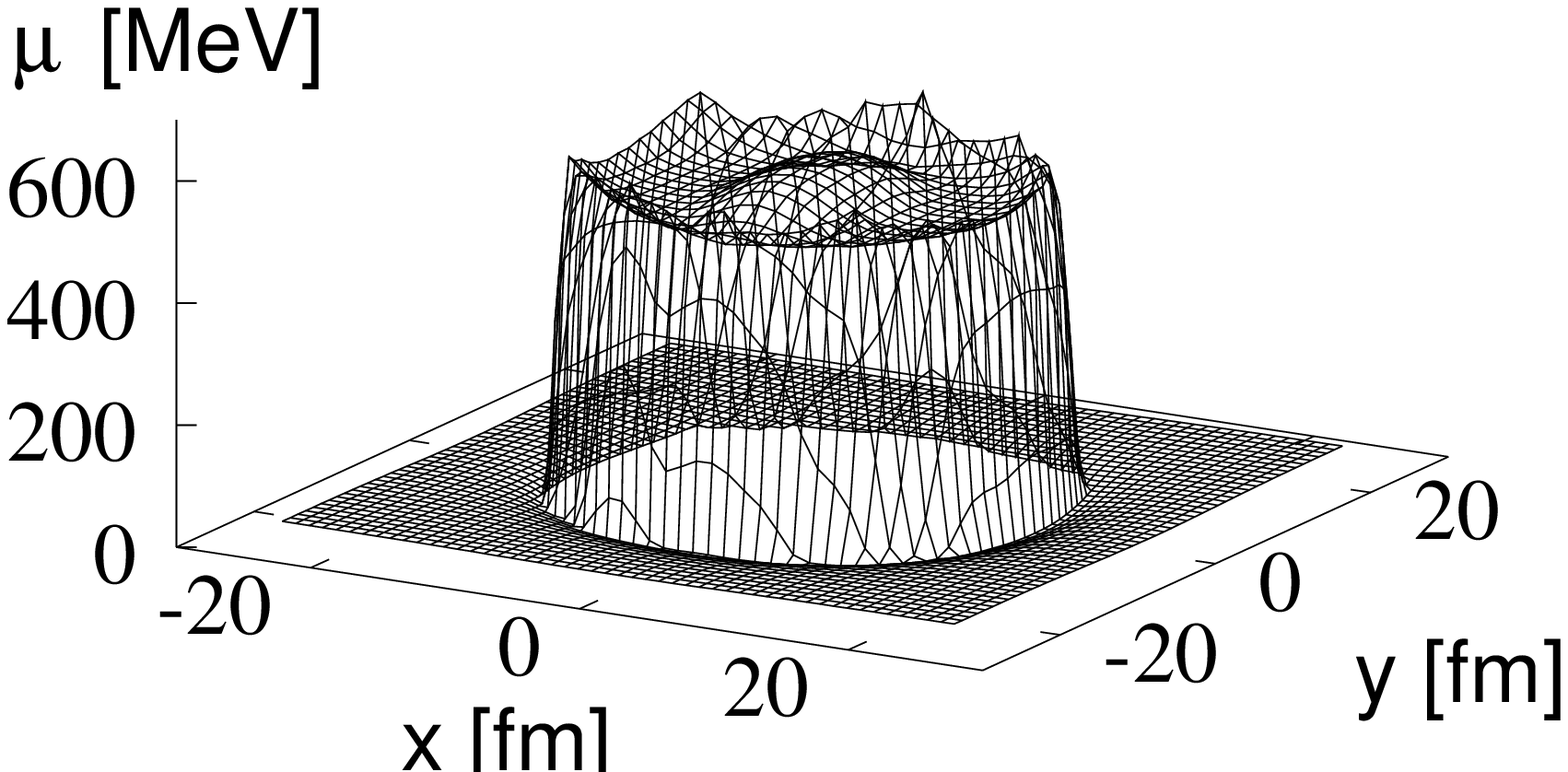,scale=0.29}
\end{minipage}
\begin{minipage}[t]{52mm}
\epsfile{file=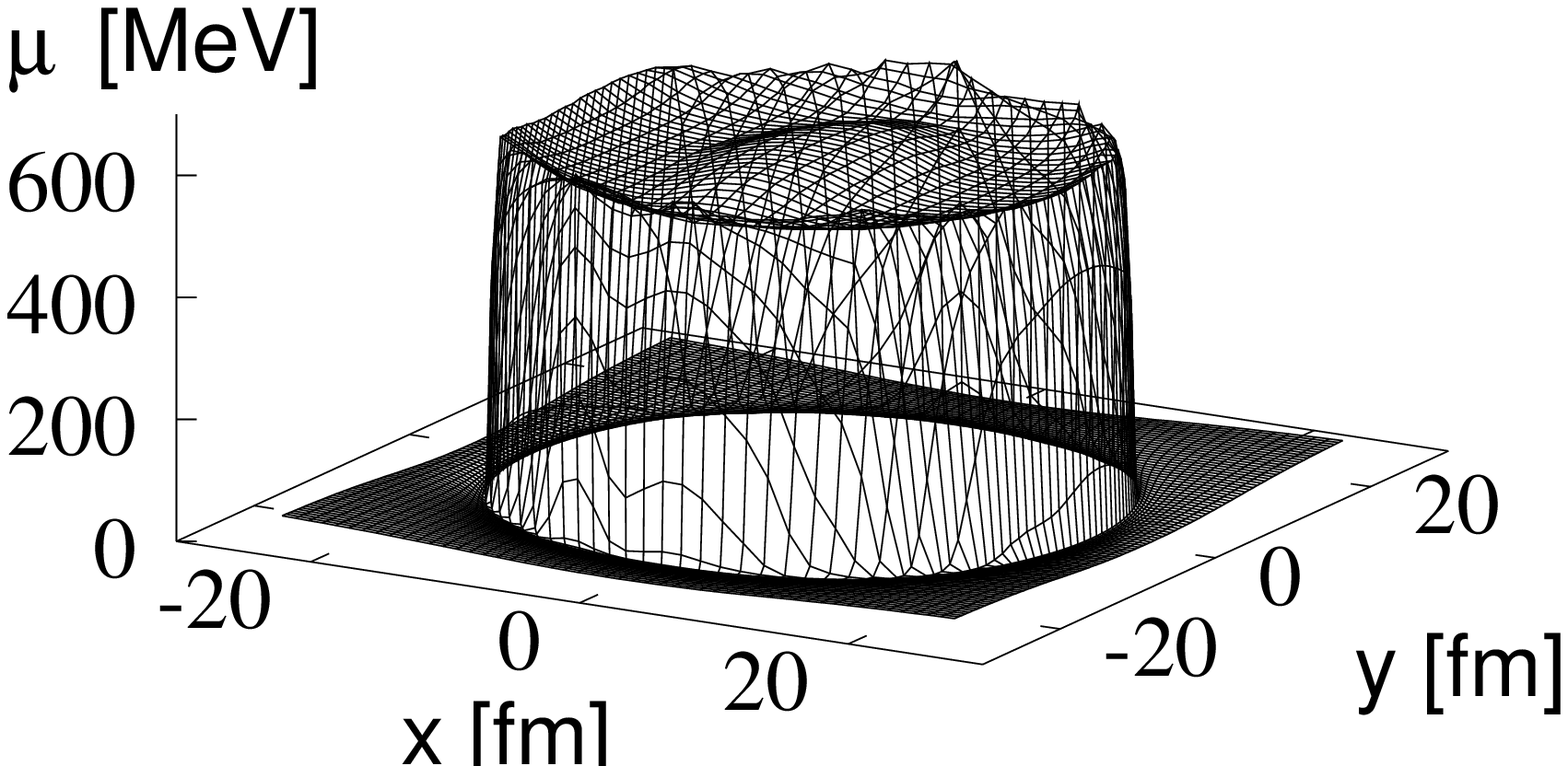,scale=0.29}
\end{minipage}
\caption{The evolution of temperature distribution  
and chemical potential distribution at $z\sim 0$ fm in Au+Au 10.8 
AGeV collision. ($b=6$ fm)}
\end{figure*}
Figure 2 displays that the evolution of temperature distribution
and chemical potential distribution at $z\sim 0$ fm in Au+Au 
10.8 AGeV collision. 
The graphs at $t=5.0$ fm/c show initial conditions from URASiMA and 
from initial time hydrodynamical expansion starts. 
We can see that both temperature distribution and chemical potential
distribution expand in the transverse direction. 
The temperature distribution decreases with time, 
on the other hand chemical potential distribution increases 
with time. This is  because baryon number density localize and the 
average of mass of baryons is larger than 1GeV. 
The fluctuations which we can see in initial chemical 
potential distribution descend with time.   
Freeze-out hypersurface which is assumed in previous section 
is illustrated with temperature distribution in fig.2. 
The life time and size of fluid depends on  
the hydrodynamical expansion and freeze-out conditions.
The size of fluid can be estimated roughly to be about $10$ fm 
at $t=15$ fm/c from 
fig. 2. This value is larger than the experimental result, $6$ fm
 which is derived by analysis of   
2-dimensional $\pi^{+}\pi^{+}$ correlations \cite{E877}.
\begin{figure*}[htb]
\begin{minipage}[t]{50mm}
\begin{center}
$t=5$ [fm/c]
\end{center}
\epsfile{file=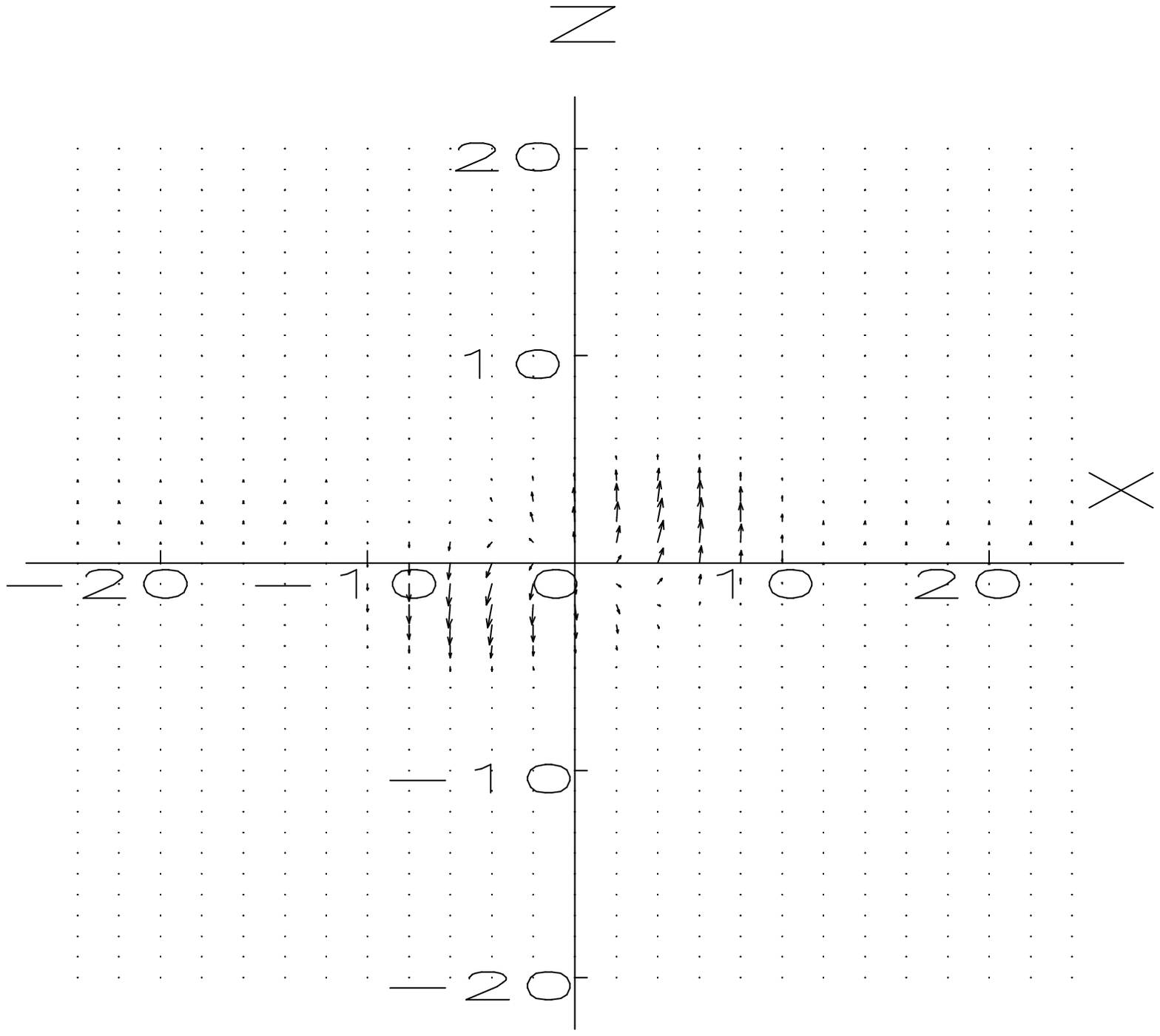,scale=0.3}
\end{minipage}
\hspace{1mm}
\begin{minipage}[t]{50mm}
\begin{center}
$t=15$ [fm/c]
\end{center}
\epsfile{file=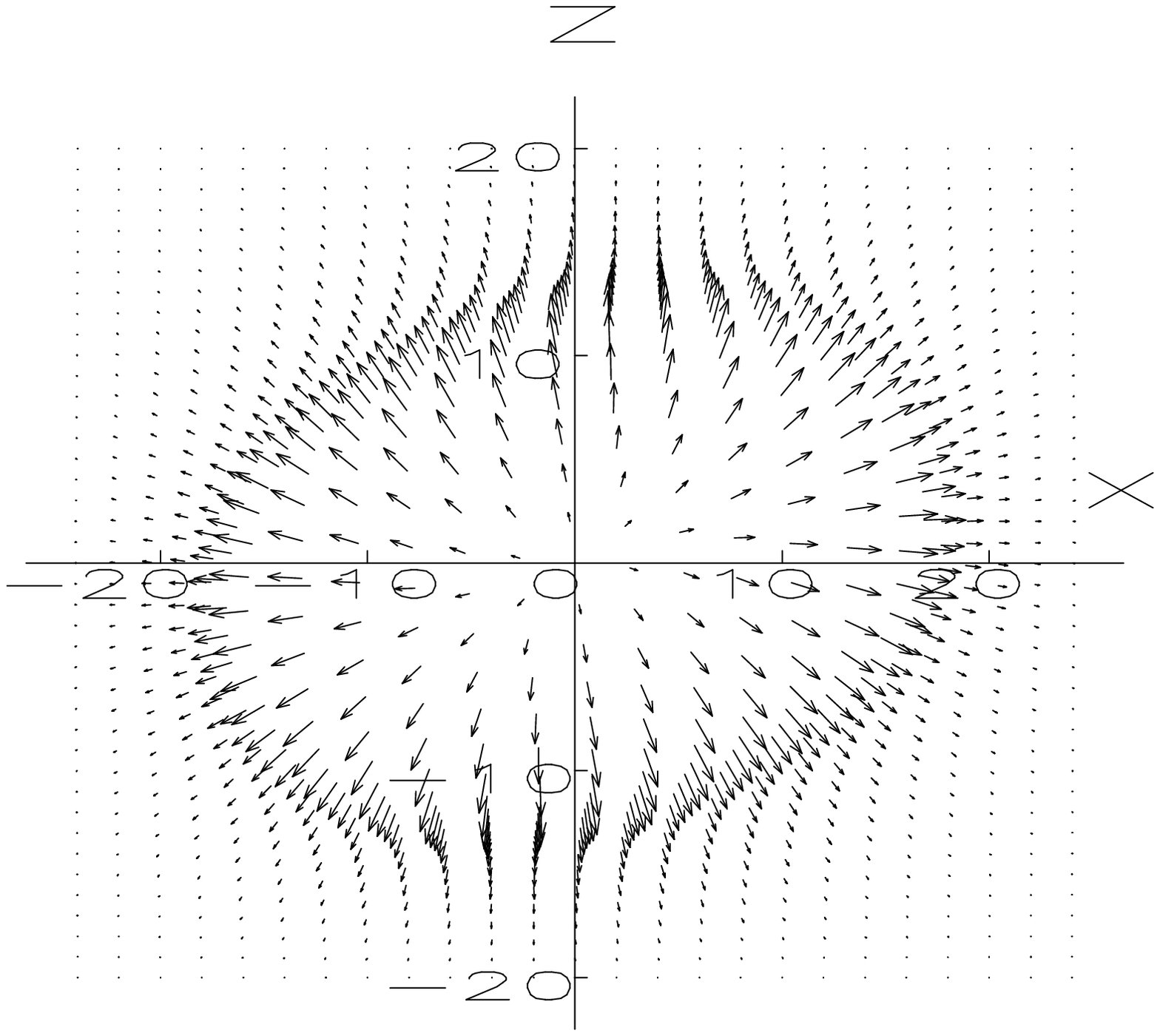,scale=0.3}
\end{minipage}
\begin{minipage}[t]{50mm}
\begin{center}
$t=25$ [fm/c]
\end{center}
\epsfile{file=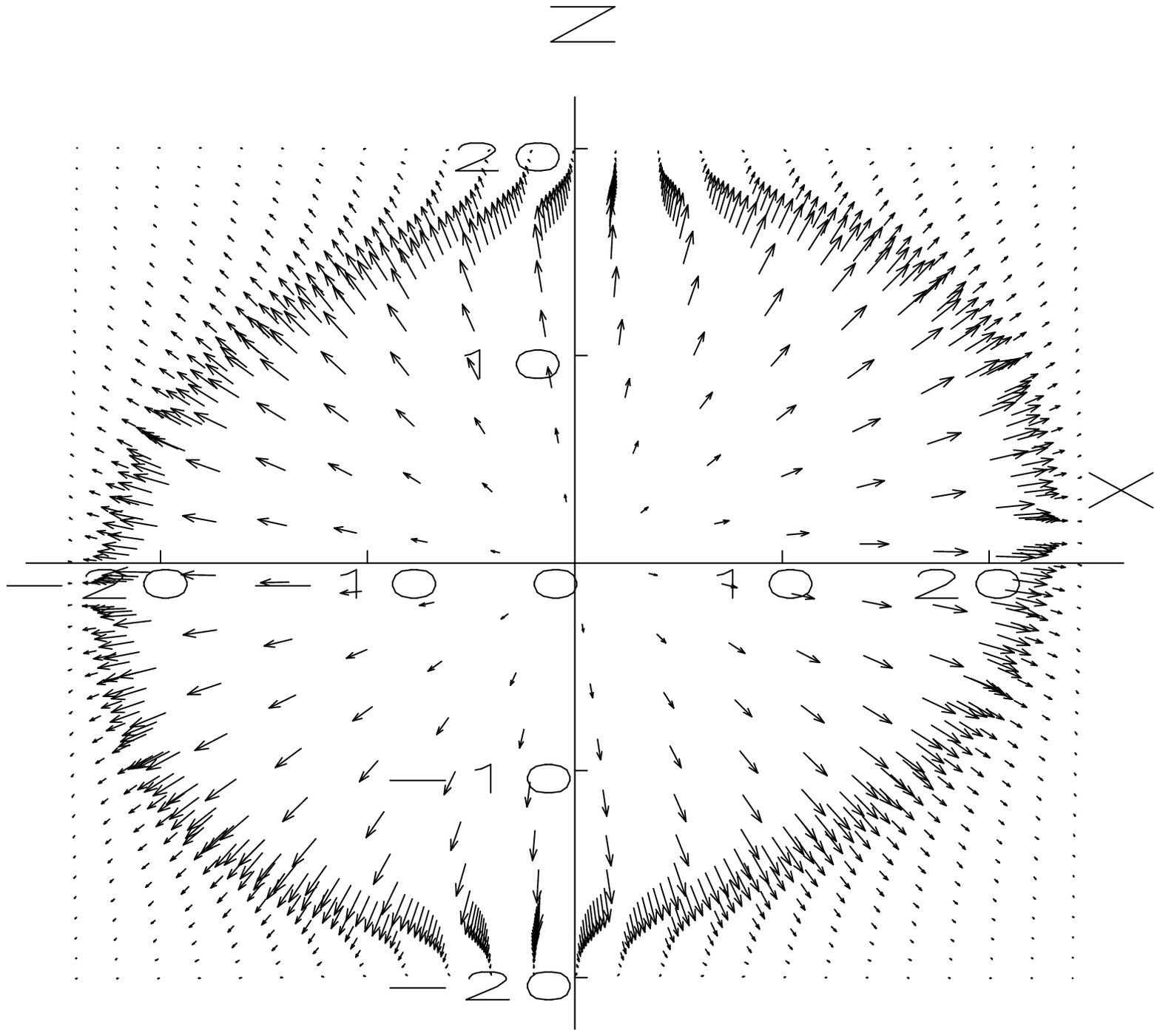,scale=0.3}
\end{minipage}
\caption{The evolution of flow at $y \sim 0$ fm in Au+
Au 10.8 AGeV collision. ($b=6.0$ fm)
}
\end{figure*}
\begin{table}[hbt]
\caption{The evolution of fluid in different incident 
energies.}
\begin{tabular*}{\textwidth}{@{}l@{\extracolsep{\fill}}ccccc}
\hline
\multicolumn{5}{l}{$t=t_{0}$ hypersurface} \\ 
\hline
$E_{\rm lab}$ [AGeV] & Max. $\epsilon$ [GeV/fm$^3$ ]& Max. $n_{\rm B}$
[fm$^{-3}$] 
& Max. $|v|$ & $t_0$ [fm/c] \\ \hline
5.0 & 1.29 & 0.79 & 0.19& 8.0 \\
10.8 & 2.06 & 0.94 & 0.28 & 5.0 \\
20.0 & 2.68 & 0.97 & 0.39 & 4.0 \\ \hline
\hline
\multicolumn{5}{l}{freeze-out hypersurface} \\ \hline
$E_{\rm lab}$ [AGeV] & Max. $\epsilon$ [GeV/fm$^3$ ]& Max. $n_{\rm B}$
[fm$^{-3}$] 
& Max. $|v|$ & Max. life time [fm/c] \\ \hline
5.0 & 0.10 & 0.083 & 0.40 & 24.0 \\
10.8 & 0.14 & 0.091 & 0.46 & 21.3 \\
20.0 & 0.16 & 0.083 & 0.48 & 21.5 \\ \hline
\end{tabular*}
\end{table}

Figure 3 shows the evolution of flow at $y \sim 0$ fm in Au+Au 10.8 AGeV 
collision.
From the initial local velocity distribution we can see 
that the projectile nuclei finishes passing though the target nuclei.
Though only small transverse flow exists at initial time, it increases 
with time and the growth of transverse flow depends on 
initial temperature distribution and chemical distribution.    
We analyze the space-time evolution of the fire ball in several initial 
conditions (table 1). 
On the $t=t_{0}$ hypersurface, maximum energy density, 
maximum baryon number density and maximum local velocity increase
with incident energy respectively.
On the freeze-out hypersurface, 
Max. $\epsilon$ and Max. $n_{B}$ are diluted, 
on the other hand Max. $|v|$ is larger than initial one.
Therefore the internal energy density turns into kinetic energy 
of the flow.
The Max. life time of fluid is not proportion to incident energy, 
because 
life time of fluid is determined not only by initial energy density 
but also by local velocity on the freeze-out hypersurface.
Figure 4 shows the $P_{\rm T}$-azimuthal fluctuation 
of proton 
in Au+Au 10.8 AGeV collision. Elliptic flow is obviously seen 
in larger $P_{\rm T}$ region. Detailed analysis will be presented 
elsewhere \cite{Nonaka}. 
 
\begin{figure*}
\begin{minipage}[t]{80mm}
\epsfile{file=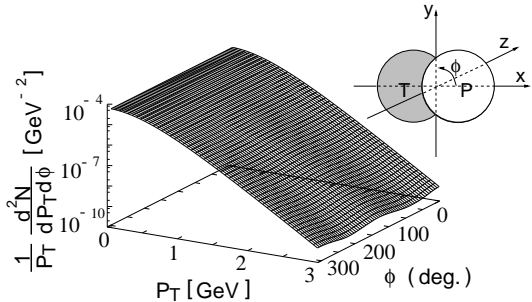,scale=0.3}
\end{minipage}
\begin{minipage}[b]{70mm}
\caption{The dependence of the transverse momentum distribution
on the azimuthal angle in Au+Au 10.8 AGeV collision. ($b=6.0$ fm)}
\end{minipage}
\end{figure*}

\section{SUMMARY}
By using full (3+1)-dimensional hydrodynamical calculation coupled 
with the baryon number conservation law,  
we investigate the hydrodynamical evolution of hot and dense matter by 
inputting the results of the event generator, URASiMA for the 
realistic initial conditions.
From analysis of the evolution of fluid with this model, 
we obtain several results at AGS energy region.
As a result of our freeze-out conditions, the life time of fluid is 
comparatively long 
 and the size of fluid which is estimated 
roughly is larger than experimental result which is 
obtained by analysis based on HBT \cite{E877}.  
By comparing physical properties on the $t=t_{0}$ hypersurface 
to ones on the freeze-out hypersurface, the decrease of energy 
density reflects on the growth of local velocity. 
Finally, from fig.4 the influence of elliptic flow to the 
particle distribution increases with transverse momentum.


\end{document}